\documentclass[12pt]{article}
\usepackage{times}
\usepackage{graphicx}
\usepackage{epstopdf}
\usepackage{doublespace}
\input epsf

\newenvironment{sciabstract}{%
\begin{quote} \bf}
{\end{quote}}

\newcounter{lastnote}

\def\wisk#1{\ifmmode{#1}\else{$#1$}\fi}

\def\lsim   {\wisk{_<\atop^{\sim}}}
\def\gsim   {\wisk{_>\atop^{\sim}}}

\def\arcmin  {\wisk{^\prime}}
\def\arcsec  {\wisk{^{\prime\prime}}}
\def\micron  {\wisk{\mu{\rm m}}}

\def\um     {$\mu$m}

\def\nwm2sr {\wisk{\rm nW/m^2/sr\ }}
\def\nw2m4sr {\wisk{\rm nW^2/m^4/sr\ }}

\title{Tracing the first stars with cosmic infrared background fluctuations}

\author{
A. Kashlinsky$^{1,{a},\ast}$, R. G. Arendt$^{1,a}$, J. Mather$^{1,{b}}$, S. H. Moseley$^{1,b}$\\
\\
\normalsize{$^{1}$Observational Cosmology Laboratory, Code 665, Goddard Space Flight Center, Greenbelt MD 20771}\\
\normalsize{$^{a}$SSAI, $^{b}$ NASA}\\
\normalsize{$^\ast$To whom correspondence should be addressed;
E-mail: kashlinsky@stars.gsfc.nasa.gov} }

\date{}
\begin{document}


\maketitle


\begin{sciabstract}
The deepest space and ground-based observations find
metal-enriched galaxies at cosmic times when the Universe was $<1$
Gyr old. These stellar populations had to be preceded by the
metal-free first stars, Population III. Recent cosmic microwave
background polarization measurements indicate that stars started
forming early when the Universe was $\lsim200$ Myr old.
Theoretically it is now thought that Population III stars were
significantly more massive than the present metal-rich stellar
populations. While such sources will not be individually
detectable by existing or planned telescopes, they would have
produced significant cosmic infrared background radiation in the
near-infrared, whose fluctuations reflect the conditions in the
primordial density field. Here we report a measurement of diffuse
flux fluctuations after removing foreground stars and galaxies.
The anisotropies exceed the instrument noise and the more local
foregrounds and can be attributed to emission from massive
Population III stars, providing observational evidence of an era
dominated by these objects.
\end{sciabstract}

The cosmic infrared background (CIB) is generated by emission from
luminous objects during the entire history of the Universe
including epochs at which discrete objects are inaccessible to
current telescopic studies \cite{hauseranddwek,kashlinsky2004}.
With new powerful telescopes individual galaxies are now found out
to redshifts of $z\gsim 5-7$,  but the period preceding that of
the galaxies seen in the Hubble Ultra-Deep Field (UDF) remains
largely unexplored. The UDF galaxies and even the highest $z$
quasars ($z\gsim 6.5$) appear to consist of ``ordinary"
metal-enriched Population I and II stars, suggesting that the
metal-free stars of Population III lived at still earlier epochs.
Large-scale polarization of the cosmic microwave background (CMB)
\cite{wmap_pol} suggests that first stars formed early at $z\sim
20$.

Current theory predicts that Population III  objects were very
massive stars with mass $> 100M_\odot$
\cite{abel,bromm,pop3review}. They should have produced a
significant diffuse background \cite{rees} with an intensity
almost independent of the details of their mass function
\cite{kagmm}. Because much of the emission longward of the
rest-frame Lyman limit from these epochs is now shifted into the
near-IR (NIR), these stars could be responsible for producing
much, or all, of the observed NIR CIB excess over that from normal
galaxy populations (see detailed review in \cite{kashlinsky2004}
and refs.
\cite{paper4,salvaterra,magliochetti,santos,cooray,kagmm}). Two
groups \cite{cooray,kagmm} have recently suggested that this
emission should have a distinct angular spectrum of anisotropies,
which can be measured if the contributions from the ordinary
(metal-rich) galaxies and foregrounds can be isolated and removed,
and provide an indication of the era made predominantly of the
massive Populations III stars.

Measuring CIB anisotropies from objects at high $z$ is difficult
because the spatial fluctuations are small and can be hidden by
the contributions of ordinary low-$z$ galaxies as well as
instrument noise and systematic errors in the data. Previous
attempts to measure the structure of the CIB in the near--IR  on
the degree scale with {\it COBE}/DIRBE \cite{paper3}, {\it
IRTS}/NIRS \cite{matsumoto} were limited in sensitivity because of
the remaining contributions from brighter galaxies in the large
beams. Analysis of 2MASS data at 1.25, 1.65 and 2.2 \um\ with
$\sim 2$ arcsec resolution  \cite{komsc,okmsc} allowed removal of
foreground galaxies to a K magnitude of 19-19.5 ($m_{AB}\sim 21$
or 15 $\mu$Jy) and led to measurements of CIB fluctuations at 1.25
to 2.2 \um\  at sub-arcminute angular scales. These studies
reported fluctuations in excess of that expected from the observed
galaxy populations, although their accurate interpretation in
terms of high-$z$ contributions is difficult due to foreground
galaxies and non-optimal angular scales.

We used data from deep exposure data obtained with {\it
Spitzer}/IRAC \cite{iraccounts,fazio_irac} in four channels
(Channels 1-4 correspond to wavelengths of 3.6, 4.5, 5.8 and 8
\um\ respectively) in attempting to uncover this signal. We find
significant CIB anisotropies after subtracting galaxies
substantially fainter than was possible in prior studies, i.e.
down to $m_{\rm AB} \sim 22-25$. The angular power spectrum of the
anisotropies is significantly different from that expected from
Solar System and Galactic sources, its amplitude is much larger
than what is expected from the remaining faint galaxies, and can
reasonably be attributed to the diffuse light from the Population
III era.

{\bf Assembly and reduction of data sets}. The primary data set
used here is the deep IRAC observation of a $\sim 12 \arcmin
\times 6 \arcmin$ region around QSO HS 1700+6416 obtained by the
Spitzer instrument team \cite{barmby,iraccounts}. In addition, the
data from two auxiliary fields with shallower exposures (HZF and
EGS) were analyzed. Relevant data characterising the fields are
listed in Table 1. For this analysis the raw data were reduced
using a least-squares self-calibration method \cite{calibration};
the processing is described in more detail in the Supplementary
Information (hereafter SI).

\begin{table}
\caption{{\bf Analyzed fields, their coordinates, exposure times,
limiting magnitudes, and pixel scales}}
\begin{center}
\begin{tabular}{c|cccc}
\hline
Region & QSO 1700 (Ch 1-4) & HZF (Ch 1-3) & HZF (Ch 4) & EGS (Ch 1-4)\\
\hline
$(\alpha, \delta)$ & (255.3, 64.2) & (136.0, 11.6) & (285.7, $-17.6$)  & (215.5, 53.3)\\
$(l, b)_{\rm Gal}$ & (94.4, 36.1) & (217.5, 34.6) & (18.4, $-10.4$) & (96.5, 58.9)\\
$(\lambda, \beta)_{\rm Ecl}$ & (194.3, 83.5) & (135.0, $-4.9$) & (285.0, 5.0) & (179.9, 60.9)\\
$\langle t_{\rm obs} \rangle (hrs)$ & 7.8, 7.8, 7.8, 9.2 & 0.5, 0.5, 0.5 & 0.7 & 1.4, 1.4, 1.4, 1.4\\
$m_{\rm Vega, lim}$ & 22.5, 20.5, 18.25, 17.5 & 21.5, 19.5, 17 & 14.5 & 22.5, 20.75, 18.5, 17.75\\
Pixel scale ($\arcsec$) & 0.6 & 1.2 & 1.2 & 1.2\\
Field size (pix) & $1,152 \times 512$ & $576 \times 256$ & $576
\times 256$ & $640 \times 384$\\\hline
\end{tabular}
\end{center}
QSO 1700 was observed during In-Orbit Checkout (IOC) with eight
AORs (Astronomical Observation Requests) which used various dither
patterns and 200-sec frame times, except for two which used
100-sec frame times (AOR ID numbers = 7127552, 7127808, 7128064,
7128320, 7128576, 7475968, 7476224, 7476480). Because of the focal
plane offset between the shared 3.6/5.8 $\micron$ and 4.5/8
$\micron$ fields of view, each of these pairs observes separate
fields which have a common overlap of $\sim 5\arcmin\times
5\arcmin$ at all four wavelengths. To provide contrast for the
self-calibration algorithm, data from a high-zodiacal light
brightness field (HZF) was co-processed for each channel. For 3.6,
4.5, and 5.8 $\micron$, the nearest suitable (200-second frame
time) data were observed later during IOC (AORID number =
8080896). For the 8 $\micron$\ data, because nominal 100 and
200-second frame times are split into pairs and quartets of
50-second frames, more nearly contemporaneous observations earlier
during IOC were used (AORID number = 6849280). For this work we
only self-calibrated the six 200-second AORs for 3.6 - 5.8
$\micron$, but at 8 $\micron$ all eight AORs were used since all
produce data with 50-second frame times. Observations of the
extended Groth Strip area (EGS) provide an additional deep data
set for verification of our results. Separate exposure times and
limiting magnitudes apply for Channels 1 - 4 respectively. The
maps for the main QSO 1700 field are shown in SI (the
supplementary information available at Nature online).
\label{table1}
\end{table}

The random noise level of the maps was computed from two subsets
(A, B) containing the odd and even numbered frames, respectively,
of the observing sequence. The A and B subsets were observed
nearly simultaneously and with similar dither patterns and
exposures. The difference between maps generated from the A and B
subsets should eliminate true celestial sources and stable
instrumental effects and reflect only the random noise of the
observations.

Analysis of the background fluctuations must be preceded by steps
that eliminate the foreground Galactic stars and the galaxies
bright enough to be individually resolved. The primary means of
removing these sources is an iterative clipping algorithm which
zeros all pixels (and a fixed number of neighboring pixels,
$N_{\rm mask}\times N_{\rm mask}$) which are more than a chosen
factor, $N_{\rm cut}$, above the 1$\sigma$ RMS variation in the
clipped surface brightness. This must be restricted to relatively
high $N_{\rm cut}$ to (a) leave enough area for a robust Fourier
analysis of the map, and (b) avoid clipping into the background
fluctuation distribution. This means that faint sources, the faint
outer portions of resolved galaxies, and the faint wings of the
point source response function around bright stars cannot be
clipped adequately. To remove these low surface brightness sources
we used a CLEAN algorithm \cite{clean} to model the entire field
in each channel. This model, convolved with the full IRAC point
spread function (PSF), was subtracted from the unclipped regions
of the map. SI provides details on this process and illustrations
of the clipped and model-subtracted images. The final step is the
fitting and removal of the 0th and 1st order components of the
background in the unclipped regions. This is done to minimize
power spectrum artifacts due to the clipping, and because the lack
of an absolute flux reference measurement and observing
constraints prohibit unambiguous determination of these
components. With this subtraction, the images represent the
fluctuation fields, $\delta F(${\mbox{\boldmath$x$}), rather than
the absolute intensity. For each observed field these steps were
carried out for images derived from the full data set and from the
A and B subsets.

{\bf Power spectrum computation and analysis.} For each channel,
we calculate the power spectra of the fluctuations as a
quantitative means of characterizing their scale and amplitude.
The power that remains after subtraction of the random noise
component is insensitive to the details of the source clipping,
and is statistically correlated between channels. This confirms
that the fluctuation signal does have a celestial origin. The
shape and amplitude of the power spectra are not consistent with
significant contributions from the cirrus of the interstellar
medium (except perhaps at 8$\mu$m) or from the zodiacal light from
local interplanetary dust.

The fluctuation field, $\delta F(${\mbox{\boldmath$x$}), was
weighted by the observation time in each pixel,
$w($\mbox{\boldmath$x$}$) \propto t_{\rm
obs}($\mbox{\boldmath$x$}), and its Fourier transform,
$f($\mbox{\boldmath$q$}$)= \int \delta
F($\mbox{\boldmath$x$}$)w($\mbox{\boldmath$x$}$)
\exp(-i$\mbox{\boldmath$x$}$\cdot$\mbox{\boldmath$q$}$) d^2x$
calculated using the fast Fourier transform. (The weighting is
necessary to minimize the noise variations across the image, but
we also performed the same analysis without it and verified that
the weight adds no structure to the resultant power spectrum. This
is because the weights are relatively flat across the image). The
power spectrum is $P_2(q)=\langle |
f($\mbox{\boldmath$q$}$)|^2\rangle$, with the average taken over
all the Fourier elements $N_q$ corresponding to the given $q$. A
typical flux fluctuation is $\sqrt{q^2P_2(q)/2\pi}$ on the angular
scale of wavelength $2 \pi/q$. In SI we show the final power
spectrum of the diffuse flux fluctuations, $P_S(q)$, and the
noise, $P_N(q)$, of the datasets. We find significant excess of
the large-scale fluctuations over the instrument noise.

Fig. \ref{fig_fin} shows the excess power spectrum,
$P_S(q)-P_N(q)$, of the diffuse light after the instrument noise
has been subtracted. There is a clear positive residual whose
power spectrum is significantly different from white noise and has
substantial correlations all the way to the largest scales probed.
Possible sources of these large-scale correlations can be
artifacts of the analysis procedure (e.g. clipping and FFT),
instrumental artifacts, local Solar System or Galactic emission,
or relatively nearby extragalactic sources and/or more distant
cosmological sources. In what follows we discuss their relative
contributions.

Residual emission from the wings of incompletely clipped sources
can give rise to spurious fluctuations, but the power spectrum of
these fluctuations should depend on the clipping parameters
$N_{\rm cut}$ and $N_{\rm mask}$. We tested the contributions from
these residuals in various ways. For a given $N_{\rm cut}$ we
varied $N_{\rm mask}$ from 3 to 7 significantly reducing any
residual wings and increasing the fraction of the clipped pixels,
but found negligible ($\lsim$ a few percent) variations in the
final $P_2(q)$. We also clipped down to progressively lower values
of $N_{\rm cut}$. For $N_{\rm cut}\lsim 3.5$ too few pixels remain
for robust Fourier analysis, so in these cases we computed
$C(\theta)$, the correlation function of the diffuse emission,
related to the power spectrum by a 1-D Legendre transformation. It
is consistent with the fluctuations in Fig. \ref{fig_fin} as
discussed and shown in SI.

Instrument noise contributions to the fluctuations were evaluated
as the power spectra of $\frac{1}{2}(\delta F_A -\delta F_B)$
using the $A-B$ subset images. As shown in SI, random instrument
noise has an approximately white spectrum. The power spectra of
the final datasets have a much larger amplitude than the noise
(especially in Channels 1 and 2) until the convolution with the
beam tapers off the signals from the sky at the smallest angular
scales. The instrument noise spectra are unaffected by the beam
and are uncorrelated from channel to channel, as expected.
However, the full data set fluctuation fields show statistically
significant correlations between channels as shown in Table 2.
This means that we see the same fluctuation field in addition to
(different) noise in all four channels. SI describes tests to
assess the contributions of possible instrumental systematic
errors. They indicate that systematic instrumental effects are
unlikely to lead to the signal shown in Fig. \ref{fig_fin}.

\begin{table}
\caption{{\bf Cross-correlation of fluctuations in units of that
of random sample ${\cal R}/\sigma_{\cal R}$ for various clipping
thresholds}}
\begin{center}
\begin{tabular}{c|cc}
\hline
Channels &  $N_{\rm cut}$ = 4 & $N_{\rm cut}$ = 2\\
\hline
1 : 2 & 52 & 12 \\
1 : 3 & 7 & 0.6 \\
1 : 4 & 10 & 4\\
\hline
\end{tabular}
\end{center}
These correlations were evaluated for an evenly covered region of
$512^2$ pixels common to all 4 channels for the QSO 1700 field.
For random uncorrelated samples, $\delta_1,\delta_2$, of $N_{\rm
pixels}$ the correlation coefficient, ${\cal R} \equiv \langle
\delta_1 \delta_2 \rangle/[\langle \delta_1^2 \rangle \langle
\delta_2^2 \rangle]^{1/2}$, should be zero with dispersion
$\sigma_{\cal R}= N_{\rm pixels}^{-1/2}$. Whereas, for $N_{\rm
mask}=3$, we find that down to even $N_{\rm cut}=2$, when only 6\%
of the pixels remain, the correlations between the channels remain
statistically significant. Simple simulations containing the
appropriate levels of the instrument noise and a power law
component of the CIB gave somewhat larger mean values of ${\cal
R}$ with the measured values lying within a 95\% confidence level;
incorporating the possibility that not all of the remaining
ordinary galaxies are the same at each wavelength would however
reduce the mean values of ${\cal R}$. \label{table2}
\end{table}

The best assessment of zodiacal light contributions to the power
spectrum comes from the examination of EGS observations taken at
two epochs 6 months apart. Because any anisotropies in the
zodiacal light cloud will not remain fixed in celestial
coordinates over this interval, the difference in the fluctuation
fields at these two epochs should eliminate Galactic and
extragalactic signals and yield a power spectrum of the zodiacal
light fluctuations added to the instrument noise and possible
systematic errors. The fluctuation levels of these difference maps
set an upper limit on the zodiacal light contribution of $<0.1$
\nwm2sr at 8 \um.  Scaling this result to the other channels by
interpolating the observed zodiacal light spectrum \cite{kelsall}
leads to zodiacal light contributions to the fluctuations that are
comfortably below the detections in Fig. \ref{fig_fin} in the
other channels as well.

Our assessment of the contribution of the infrared cirrus (i.e.
interstellar clouds of neutral gas and dust) to the power spectra
is derived from the 8 $\mu$m HZF field which lies at low $b_{\rm
Gal}$ and is visibly contaminated by cirrus. Assuming that the
large scale fluctuations in this field are due to the cirrus,
relative fluctuations of the 8 $\mu$m cirrus are $\sim1$\% of the
mean cirrus flux level. A similar level of relative cirrus
fluctuations in the QSO 1700 field would have an amplitude of
$\sim0.3$ nW/m$^2$/sr. This is not significantly lower than the 8
\um\ amplitude observed in Fig. \ref{fig_fin}. Therefore, we
cannot presently eliminate the possibility that the fluctuations
at 8 $\mu$m are dominated by cirrus. However, the spectrum of the
ISM emission should drop sharply at shorter wavelengths as the PAH
emission bands that dominate at 8 $\mu$m become less significant.
Given this estimate of the cirrus contribution at 8 \um\ we
estimate the amplitude of large-scale fluctuations for channels 1
-- 3 as $\simeq$ 0.03, 0.03, and 0.08 nW/m$^2$/sr, which are well
below the observed fluctuations in these channels.

Colour is another important criterion for testing the origins of
the signal in Fig. \ref{fig_fin}. The cross-correlations between
the channels for the common area of the QSO 1700 field are
statistically significant and they also strengthen at the larger
scales where the noise contribution is smaller, as verified by
smoothing the maps. These significant cross-correlations allow us
to examine the corresponding colours of the fluctuations. We made
several estimates of the colours, $\beta_{n1}$, between channels
$n$ and 1: a) as the square root of ratio of the power spectra,
and b) as $\beta_{n1}(\theta) = \langle
\delta_n(x)\delta_1(x+\theta)\rangle/\langle
\delta_1(x)\delta_1(x+\theta)\rangle$ evaluated over the 512$^2$
pixel field common to all four channels, where a further
consistency check comes from comparison between $\beta_{21}$ and
$\beta_{41}/\beta_{42}$. For channels 1 and 2 these estimates are
shown in Fig. \ref{corfun} and appear roughly independent of
angular scale and mutually consistent. The instrument noise is too
large to enable robust colour estimates involving channel 3 (5.8
$\mu$m).  These derived colours indicate that the fluctuation
signal in Fig. \ref{fig_fin} has an energy distribution that is
approximately flat to slowly rising with wavelength in $\nu
I_\nu$. The energy spectrum rules out contributions from remaining
Galactic stars, but probably cannot be used to distinguish between
ordinary galaxies and Pop III objects without additional detailed
modeling of both sets of sources.

{\bf CIB fluctuations from extragalactic sources}. There are two
extragalactic classes of contributors to CIB fluctuations:
``ordinary" galaxies containing normal stellar Populations I and
II, and the objects that preceded them, Population III stars. The
square of the CIB fluctuation in band $\nu$ produced by
cosmological sources that existed over time period $\Delta t$ is
given by the power-spectrum version of the Limber equation (see
\cite{kashlinsky2004} and SI):
\begin{equation} \frac{q^2P_2(q)}{2\pi} =
\Delta t \int \left( \frac{d I_{\nu^\prime}}{dt}\right)^2
\Delta^2(qd_A^{-1};z) dt \label{limber_del2}
\end{equation}
where $\nu^\prime=\nu(1+z)$ is the rest frequency of the emitters,
$d_A$ is the comoving angular diameter distance and
\begin{equation}
\Delta (k) = \sqrt{\frac{1}{2\pi} \frac{k^2 P_3(k)}{c\Delta t} }
\label{del_cib}
\end{equation}
 is the fluctuation in the number of sources within a volume
$k^{-2} c\Delta t$. The fluctuation of the CIB on angular scale
$\simeq 2\pi/q$ can then be expressed as $\delta F_{\rm CIB} =
F_{\rm CIB} \Delta(qd_A^{-1}(\langle z\rangle))$, where $\langle
z\rangle$ is the suitably averaged {\it effective} redshift. The
relative CIB fluctuation is a suitably averaged fluctuation in the
source counts over a cylinder of radius $q^{-1} d_A(\langle z
\rangle)$ and length $c\Delta t$. Three things would lead to
larger fluctuations: 1) a population that, after removing
constituents brighter than some limit, leaves a substantial mean
CIB flux (increase $F_{\rm CIB}$), 2) populations that existed for
a shorter time (increase $\langle\Delta\rangle$ by decreasing
$\Delta t$), and 3) populations that formed out of rare peaks of
the underlying density field leading to biased and significantly
amplified, \cite{kaiser} clustering properties (increase
 $\langle\Delta\rangle$ by increasing $P_3(k))$.

{\bf CIB fluctuations from remaining ordinary galaxies}. The
fluctuations produced by ordinary galaxies contain two components:
1) shot noise from discrete galaxies and 2) galaxy clustering from
the primeval density field. The amplitude of the first component
can be estimated directly from galaxy counts. In order to estimate
the contribution from the second component we proceed as follows:
from galaxy count data we estimate the total CIB flux produced by
the remaining galaxies fainter than our clipping threshold.
Ordinary galaxies occupy an era of $\Delta t\gsim$ a few Gyrs.
Their present-day clustering pattern on the relevant scales is
well measured today. The clustering pattern evolution can be
extrapolated to earlier times assuming the ``concordance"
$\Lambda$CDM model. These parameters (flux, $\Delta t$, clustering
pattern and its evolution) then allow us to estimate the
contribution to the CIB fluctuations via eqs.
\ref{limber_del2},\ref{del_cib}. Because galaxy clustering is
weaker at earlier times, an upper limit on the contribution from
ordinary galaxies is obtained assuming that the clustering at
early times remained the same as at $z=0$.

The shot noise component contributed by the ordinary galaxies to
the CIB angular spectrum was estimated directly from galaxy count
data: each magnitude bin $\Delta m$ with galaxies of flux $f(m)$
would contribute a power spectrum of $P_{\rm
sn}=f^2(m)\frac{dN_{\rm gal}}{dm}\Delta m$. Fig. \ref{fig_fin}
shows the shot noise component convolved with the IRAC beam for
the galaxies at the limiting magnitude in Table \ref{table1}. The
limiting magnitudes are qualitative estimates of where the
SExtractor \cite{sextractor} number counts (in 0.5 magnitude bins)
begin to drop sharply due to incompleteness. The shot noise fits
the observed diffuse light fluctuations well at small angles, but
the shot-noise contribution from ordinary galaxies cannot make a
substantial contribution to the large scale power of the diffuse
light in Fig. \ref{fig_fin}.

We assume the ``concordance" $\Lambda$CDM model Universe with flat
geometry ($\Omega_{\rm total}$=1) dominated by a cosmological
constant, $\Omega_\Lambda\simeq 0.7$, with the rest coming from
cold dark matter and ordinary baryons in proportions suggested by
measurements of the CMB anisotropies \cite{wmap} and high-$z$
supernovae \cite{sn1a}. With $P_3(k)$ taken from the $\Lambda$CDM
concordance model, the power in fluctuations from the clustering
of ordinary galaxies depends on the net flux produced by the
remaining ordinary galaxies (via $F_{\rm CIB}$) and their typical
redshifts (via $\Delta(qd_a^{-1})$). The total flux from galaxies
fainter than the limiting magnitude in Table 1 was estimated from
the Spitzer IRAC galaxy counts (\cite{iraccounts}) and is shown in
Fig. \ref{cib_gals}. The flux contributed to the CIB from the
remaining galaxies is between 0.1 and 0.2 \nwm2sr and this
amplitude is much less than the excess CIB at these wavelengths
(\cite{kashlinsky2004}) indicating that the excess CIB flux is
produced by populations that are still farther out than the
ordinary galaxies remaining in the data. This (low) value of the
remaining flux can also be derived from the small amplitude of the
residual shot-noise contribution to the fluctuations in the
confusion-limited datasets. Thus, in order to explain the
amplitudes shown in Fig. \ref{fig_fin}, the remaining ordinary
galaxy populations would need to produce relative flux
fluctuations of order $\gsim 100\%$ from their clustering. Without
follow-up spectroscopy it is difficult to determine with high
precision the range of redshifts of the remaining ordinary
galaxies, but approximate estimates can be made and are summarized
in the Figure \ref{cib_gals} legend to be $z\gsim 1$.

In flat cosmology with $\Omega_\Lambda=0.7$ one arcminute subtends
a comoving scale between 0.7 and 1.6 $h^{-1}$Mpc at $z$ between 1
and 5. The present day 3-D power spectrum of galaxy clustering,
$P_{3,\rm gal}(k)$, is described well by the concordance
$\Lambda$CDM model (\cite{efstathiou}, \cite{tegmark}) and one
expects that on arcminute scales the density field was in the
linear to quasi-linear regime at the redshifts probed by the
remaining galaxies between 3.6 and 8 \um. At smaller scales
non-linear corrections to evolution were computed following
(\cite{peacockanddodds}). The resultant CIB fluctuation from the
remaining ordinary galaxies, producing $F_{\rm CIB, og}=0.14$
\nwm2sr , times $\Delta(\propto (\Delta t)^{-1/2})$ is shown in
Fig. \ref{fig_fin} for $\langle z\rangle$=1,3,5 and $\Delta t $=5
Gyr corresponding to the age of the Universe at $z\sim 1$. For the
$\Lambda$CDM model the relative fluctuations in the CIB  on
arcminute scales would be of order $\langle \Delta \rangle \sim
(2-10)\times 10^{-2} (\Delta t/{\rm 5 Gyr})^{-0.5}$ from galaxies
with $\langle z \rangle $=1-5 assuming no biasing. Combining this
with the above values for the diffuse flux from the remaining
ordinary galaxies would lead to $\delta F \lsim (1-2)\times
10^{-2}$ \nwm2sr in all the channels. While biasing may increase
the relative fluctuations, with reasonable bias factors for
galaxies lying at $z\sim $ (a few), the diffuse light fluctuations
would still be very small compared to those in Fig. \ref{fig_fin}
and are unlikely to account for fluctuations of amplitude $\sim
(0.1-0.5)$ \nwm2sr at arcminute scales. An upper limit on the CIB
fluctuations can be evaluated assuming the same clustering pattern
for the remaining galaxies as at the present epoch, $z=0$, i.e.
that their 2-point correlation function is given by
$\xi=(r/r_*)^{-1.7}$ with $r_*=5.5 h^{-1}$Mpc \cite{apm}; its
$\Delta$ times $F_{\rm CIB, og}$ is also shown in
Fig.\ref{fig_fin} and is much below the signal we measure.
Conversely, one can evaluate the clustering strength needed to
account for the observed fluctuations: a 100\% relative
fluctuation from galaxies at $z\gsim 1$ clustered with
$\xi=(r/r_*)^{-1.7}$ would require $r_* \gsim 25 h^{-1}$Mpc. This
corresponds to the effective bias factor $\gsim 5$ which is
significantly higher than expected from gravitational clustering
evolution in the $\Lambda$CDM Universe \cite{springel}. In fact,
direct measurements of clustering for relatively nearby IRAC
galaxies with flux $>32 \mu$Jy at 3.6 \um\ ($\sim$5 magnitudes
brighter than the limit reached for remaining galaxies in the QSO
1700 field) \cite{swire} find the projected 2-point correlation
amplitude on $\sim$arcmin scales of $w_{>32\mu{\rm
Jy}}(\theta\gsim 1^\prime)<4\times 10^{-2}$ corresponding to
relative fluctuations in CIB from these galaxies of amplitude
$\sim \sqrt{w}\lsim 0.2$. Fainter galaxies are expected to have an
even lower correlation amplitude.

The QSO 1700 field seems to contain an overdensity of Lyman-break
galaxies at $z\simeq 2.3$ \cite{steidel}, which could lead to a
larger $\Delta$ and CIB fluctuations for {\it this} region.
However, we see similar levels of fluctuations for the other
fields located at very different parts of the sky making it
unlikely that the overdensity claimed in ref. \cite{steidel} can
account for our signal. When account is made of the different shot
noise levels from the remaining galaxies, the fluctuations seen in
the different fields have consistent power spectra within the
statistical uncertainties. (An exception being channel 4 HZF data,
located at low $b_{\rm Gal}$, and clearly dominated by Galactic
cirrus).

{\bf  CIB fluctuations from the Population III era}. Population
III at $z\sim$10-30 are expected to precede ordinary galaxy
populations. One can expect on fairly general grounds \cite{kagmm}
that, if massive, they would contribute significantly to the NIR
CIB, both its mean level and anisotropies. Intuitive reasons are
discussed in \cite{kashlinsky2004}, but as eqs.
\ref{limber_del2},\ref{del_cib} shows they are mostly related to:
1) if massive, Population III were very efficient light emitters,
2) their era likely lasted a shorter time, $\Delta t$, than that
of the ordinary galaxies, leading to larger $\Delta$ in
eq.\ref{del_cib}, and 3) they should have formed out of high peaks
of the density field whose correlation function is strongly
amplified. The near-IR also probes UV to visible parts of the
electromagnetic spectrum at $z\sim 10-30$, where most of their
emission is produced \cite{santos}.

The amplitude of the CIB anisotropies remaining in the present
data implies that the remaining CIB originates from still fainter
objects. Can the observed amplitudes of fluctuations in Fig.
\ref{fig_fin} be accounted for by energetic sources at high $z$?
Population III stars can produce significant NIR CIB levels
\cite{santos}, $\gsim 1$ \nwm2sr and e.g. the NIR CIB excess over
that from ordinary galaxies at 3.6 \um\ is $8.7 \pm 3.1$ \nwm2sr
\cite{kashlinsky2004}. Thus Population III objects would require
smaller relative CIB fluctuations. Because individual Population
III systems are small, yet numerous, the shot-noise component of
the CIB from them is small and, in any case, is already absorbed
in the shot-noise shown in Fig. \ref{fig_fin}. Additionally, their
$\langle\Delta\rangle$ would be amplified by the much shorter
$\Delta t$ and (significant) biasing, and Population III
contribution would dominate the diffuse light fluctuations in Fig.
\ref{fig_fin}. Detailed theoretical interpretation of the results
in terms of the Population III era models exceeds the scope of
this article, but qualitative comparison can be made by estimating
the typical value of the relative CIB fluctuation, $\Delta$,
corresponding to that era. The fraction of the Population III
haloes was calculated assuming that they form from the
$\Lambda$CDM density field in haloes where the virial temperature
$T_{\rm vir} \geq 2000$ K to enable efficient molecular hydrogen
cooling (\cite{miralda}). Biasing was treated using the
gravitational clustering prescription from \cite{kashlinsky1998}
in the $\Lambda$CDM model. (Non-linear evolution effects are small
on the angular scales and redshifts probed here.) Assuming $\Delta
t$=300 Myr, which corresponds to the age of the Universe at
$z$=10, we get typical values of $\Delta \simeq$ 0.1-0.2 at $z$
between 10 and 20. This order-of-magnitude evaluation shows that
the levels of $\sim$0.1-0.3 \nwm2sr at 3.6 to 8 \um\ on arcminute
scales can be accounted for by Population III emissions if their
total flux contribution is $> 1$ \nwm2sr which is reasonable as
discussed earlier.

Earlier studies of CIB fluctuations
\cite{paper3},\cite{komsc},\cite{matsumoto} contained significant
contributions from relatively bright galaxies making it difficult
to isolate the possible CIB fluctuations from the very early
times. The contribution to CIB fluctuations from remaining
galaxies is a function of the limiting magnitude below which
galaxies are removed. With the Spitzer IRAC data we could identify
and remove galaxies to very faint limits of flux $\gsim0.3 \mu$Jy.
This limit is, at last, sufficiently low to push the residual
contribution from ordinary galaxies along the line-of-sight below
the level of the excess signal at larger angular scales. If our
interpretation is correct and the signal we detect comes from
Population III located at much higher $z$, the amplitude of the
CIB fluctuations on scales where galaxy shot-noise is negligible
should remain the same as fainter ordinary galaxies are removed
with deeper clipping. This is true as far as we can test with this
data and would certainly be verifiable with longer exposure data.

At these $z$ the IRAC bands probe the rest-frame wavelengths
between 0.2 and 0.8 \um, where the energy spectrum of individual
Population III object is dominated by free-free emission and is a
slowly rising function of wavelength \cite{santos}; this would be
consistent with the spectrum of the CIB anisotropies in Fig.
\ref{fig_fin}. Near-IR observations at shorter wavelengths would
be particularly important in confirming the redshifts where the
CIB fluctuations originate as there should be a significant drop
in the fluctuations power at the rest-frame Lyman limit
wavelength.

{\bf Acknowledgements} We thank Giovanni Fazio for access to the
IRAC Deep Survey data and Dale Fixsen and Gary Hinshaw for
comments on drafts of this paper. This material is based upon work
supported by the National Science Foundation. This work is based
on observations made with the Spitzer Space Telescope, which is
operated by the Jet Propulsion Laboratory, California Institute of
Technology under a contract with NASA. Support for this work was
also provided by NASA through an award issued by JPL/Caltech.

{\bf Competing interests statement}: The authors declare that they
have no competing financial interests.

{\bf Correspondence} and requests for materials should be
addressed to AK (kashlinsky@stars.gsfc.nasa.gov).

{\bf Author contributions}: AK is responsible for the idea,
clipping the maps, power spectrum and correlation analyses,
evaluating the extragalactic contributions and  writing the paper.
R.G.A. is responsible for the images for analysis, providing the
model of the resolved sources with the IRAC PSF, and evaluating
systematics, instrument and zodiacal and cirrus contributions. JM
and SHM developed analysis strategy and searched for alternate
explanations for the fluctuations. All authors provided critical
review of the analysis techniques, results, and manuscript.

\newpage

\begin{figure}[h]
\centering \leavevmode \epsfxsize=1.05 \columnwidth
\includegraphics[width=7in]{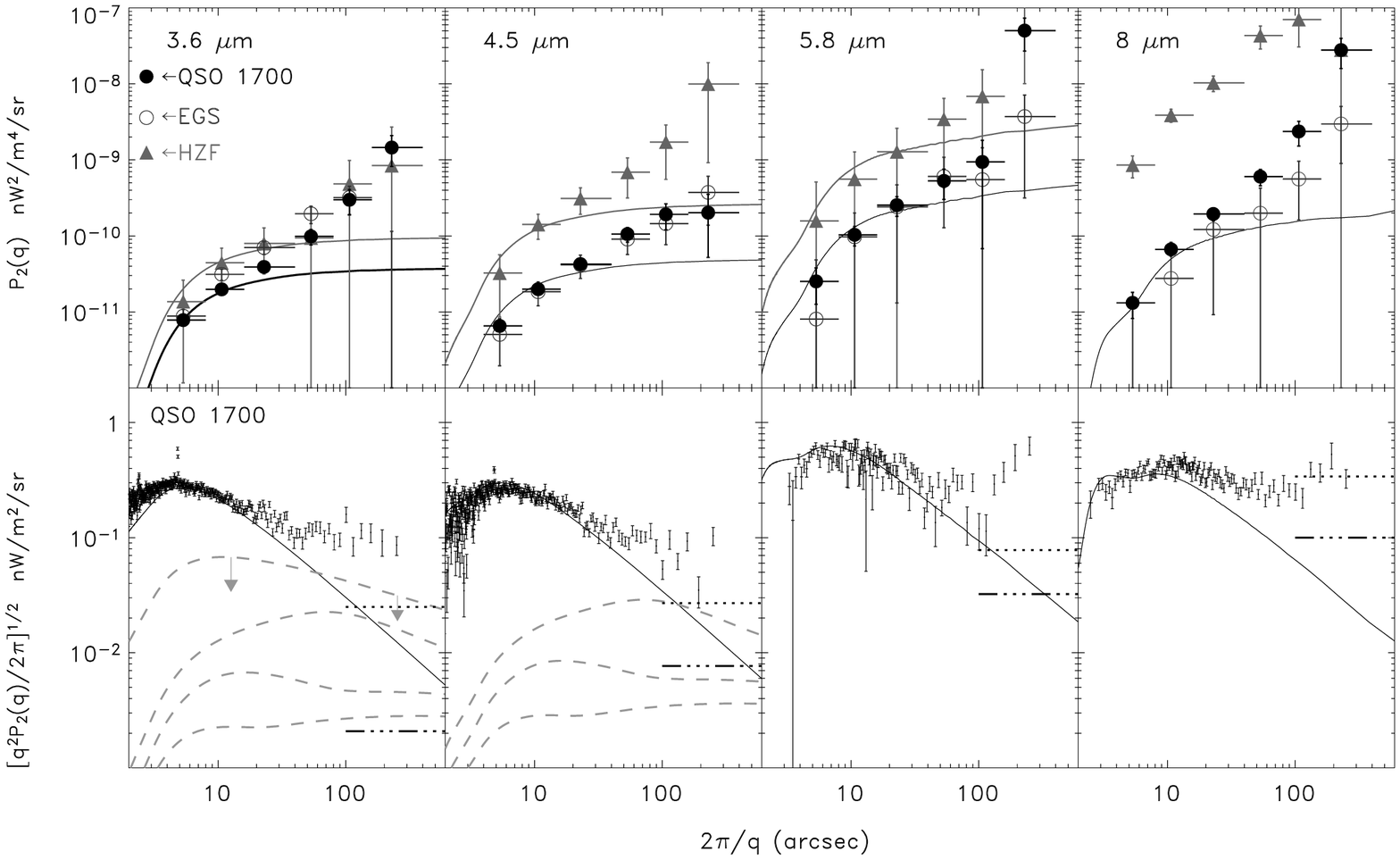} \vspace{0.5cm} \caption[]{{\bf Spectrum of CIB
fluctuations} Top: Power spectrum of signal minus noise from the
SI Fig. 2 averaged over wide bins to increase signal-to-noise. The
errors are $N_q^{-1/2}$ corresponding to the cosmic variance,
where $N_q$ is number of Fourier elements at the given $q$-bin.
Solid lines show the shot noise from remaining galaxies fainter
than the limiting magnitude in Table 1. Filled circles and the
darkest shade error bars correspond to the QSO1700 data, open
circles and intermediate shade error bars to the EGS data and
triangles with the lightest shade error bars and lines correspond
to HZF data. Bottom panels: fluctuations,
$[\frac{q^2P(q)}{2\pi}]^{1/2}$, vs $2\pi/q$ for the QSO1700 data.
Dashed lines estimate the contribution from ordinary galaxy
populations and $\Lambda$CDM density field with $\Delta t$=5Gyr:
the top dashed line shows the {\it upper} limit which assumes that
their (high-$z$) clustering pattern remained identical to that
today at $z$=0 and the other dashed lines correspond to $\langle
z\rangle$=1,3,5 from top to bottom. Solid lines show shot noise
from remaining ordinary galaxies. Dotted and dash-dot-dotted lines
show the estimated Galactic cirrus and zodiacal light
contributions respectively assuming the power spectra to be $P(q)
\propto q^n$ with $n=-2$ typical of cloud distributions. The
observed cirrus power spectrum is a little steeper with $n\sim
-2.5$ to $-3$ \cite{gautier,ingalls,paper3,wright1998} in which
case the lines will have a slope of $(3+n)/2$=0.25 to 0.5.}
\label{fig_fin}
\end{figure}

\newpage

\begin{figure}[h]
\centering \leavevmode \epsfxsize=0.6 \columnwidth
\includegraphics[width=5in]{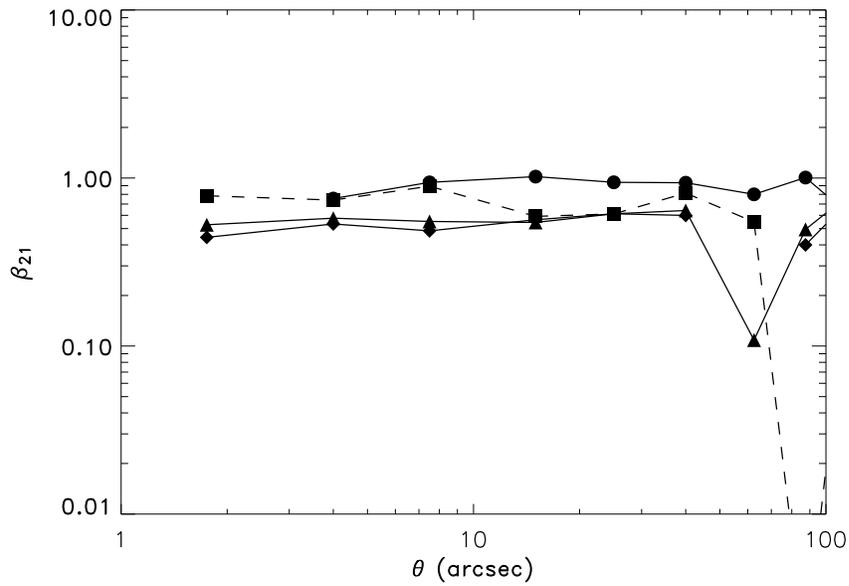} \vspace{0.5cm} \caption[]{{\bf Colour properties
of clipped maps}  Estimates of colour between channels 1 and 2.
Circles correspond to $\sqrt{P_2/P_1}$, averaged over the bins
centered at these angles, diamonds to $\langle
\delta_2(x)\delta_1(x+\theta)\rangle/\langle\delta_1(x)\delta_1(x+\theta)\rangle$
for $N_{\rm cut}=4$ and triangles to the same quantity averaged
over maps with $N_{\rm cut}=4,3,2.5,2$. Squares correspond to
$\beta_{41}/\beta_{42}$ averaged over maps with $N_{\rm
cut}=4,3,2.5,2$.}\label{corfun}
\end{figure}
\newpage

\begin{figure}[h]
\centering \leavevmode \epsfxsize=0.95 \columnwidth
\includegraphics[width=7in]{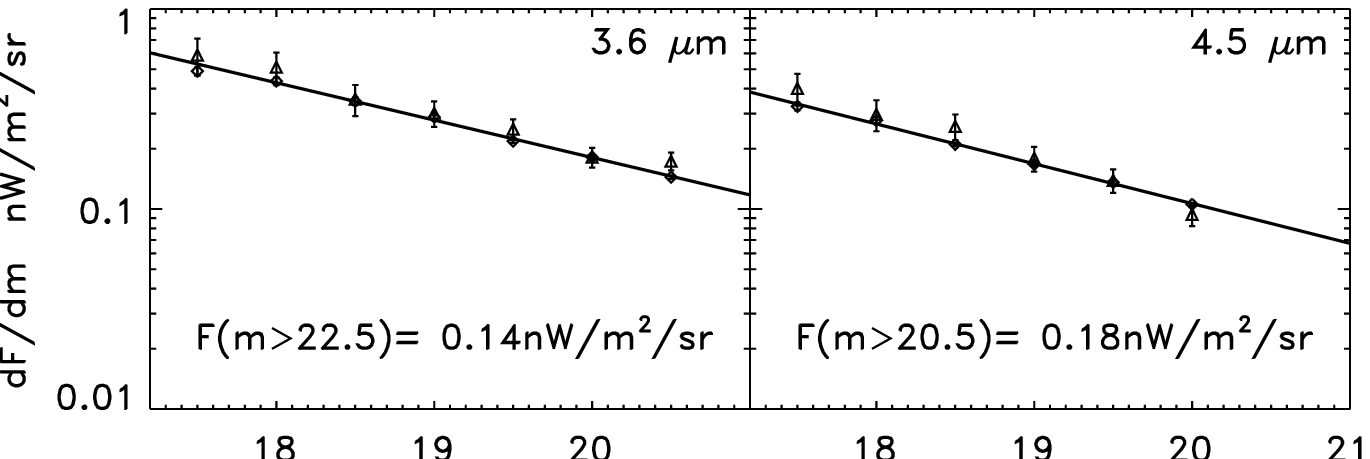} \vspace{0.5cm} \caption[]{{\bf Contribution to
CIB flux from Spitzer IRAC galaxy counts at 3.6 \um\ and 4.5 \um.}
The value of $dF/dm$ is evaluated from counts data and is a
rapidly decreasing function of $m$ of the form $dF/dM \propto
\exp(-bm)$ with $b\simeq 0.4$ (the least squares fits are shown
with solid lines). (At the other two IRAC channels the uncertainty
in the fall-off is much greater, but we show below that definite
conclusions can already be reached from the data at 3.6 and 4.5
\um). Assuming that the functional form of $dF/dm$ does not begin
to rise appreciably at still fainter magnitudes gives the CIB
contribution from galaxies fainter than $m_0$ of
$F(m>m_0)=b^{-1}dF/dm|_{m_0}$.  For the QSO 1700 field, we
eliminate galaxies brighter than $m_{\rm Vega}\simeq 22.5$ at 3.6
\um\ and $m_{\rm Vega}\simeq 20.5$ at 4.5 \um\ corresponding to AB
magnitudes $\simeq 25$ and $24$ respectively. The annotation in
each panel show numbers which correspond to the extrapolated total
flux from galaxies fainter than the magnitude limits in Table 1
using the least-squares fits shown with solid lines. One can
expect that galactic spectra at the appropriate range of
wavelengths are at most as steep as that of Vega, a star with a
Rayleigh-Jeans black body spectral fall-off at these wavelengths.
If so, they would generally have $K-m_{3.6}>0$ leaving galaxies
with $K\gsim 22$ (e.g. \cite{eisenhardt}). Galaxies at K$> 20$
have median redshift $\gsim 1$ (\cite{cowie}), so the above
argument would place the galaxies remaining at 4.5 \um\ at $z\gsim
1$ and those remaining at 3.6 \um\ still farther out to higher
$z$. Comparison with the Lyman Break Galaxy candidates in the same
area shows that a substantial fraction of these galaxies are at
$z\sim 2-4$ with $z\sim 3$ being the median redshift
\cite{barmby}. Similarly, a substantial part of these galaxies are
confirmed to be star-forming systems at $z\simeq 2.3$
(\cite{shapley}).} \label{cib_gals}
\end{figure}

\newpage
\title{\bf{Supplementary Information for ``Tracing the first stars with cosmic infrared background fluctuations"}}

\author{
by A. Kashlinsky, R. G. Arendt, J. Mather, S. H. Moseley}

\maketitle


\begin{sciabstract}
This material presents technical details to support the discussion
in the main paper. We discuss here data assembly, removal of
individual resolved sources, computations of the power spectra and
correlation functions from the maps cleaned off these sources, and
evaluation of the Limber equation in the form used in the main
paper. We also show maps of the cleaned field and present the
plots with their power spectrum and that of the instrument noise.
\end{sciabstract}

{\bf Data processing}

For this analysis the raw data were reduced using a least-squares
self-calibration method (ref. [18] in main paper). This procedure
models the raw data ($D^i$) of each frame as
\begin{equation}
D^i = G^p S^{\alpha} + F^p + F^q
\renewcommand{\theequation}{SI-\arabic{equation}}
\setcounter{equation}{1}
\end{equation}
where the index $i$ counts over all pixels of all frames,
$S^{\alpha}$ is the sky intensity at location $\alpha$, $G^p$ is
the detector gain at pixel $p$, $F^p$ is the detector offset at
pixel $p$, and $F^q$ is an offset that can differ for each frame
and each of the four detector readouts (combined into the index
$q$). This self-calibration procedure has advantages over the
standard pipeline calibration of the data in that the derived
detector gain and offsets match the detector at the time of the
observation, rather than at the time of the calibration
observations. The most noticeable difference (especially at 3.6
and 8 $\micron$) is that the different AORs are affected by
different patterns of residual images left by prior observations,
which if not properly removed, can leave artifacts in the final
images.

The combined use of low (QSO 1700) and high brightness (HZF)
fields removes the gain-offset degeneracy in the self-calibration,
although since we are interested in the nearly flat uniform
background, to first order it is more important that the observing
strategy included dithering to distinguish between detector and
sky variations. Because IRAC's calibration does not include
zero-flux closed-shutter data, a degeneracy remains in the
absolute zero point of the solution. The sky intensity
$S^{\alpha}$ can be increased by a constant value at all locations
$\alpha$, and the offset $F^p$ can be appropriately decreased
without affecting the quality of the fit. A degeneracy also
remains in first-order gradients (or two-dimensional linear
backgrounds). The parameter $F^q$ can correlate with the dither
position to induce false linear gradients in the other parameters
without affecting the quality of the fit. However higher-order
gradients are not degenerate. Because of these degeneracies,
linear gradients have been fit and subtracted from the
self-calibrated mosaicked images. Hence our estimates of the power
at scales $\gsim 5^\prime$ should be treated as lower limits.

After the self-calibrations had been determined, corrections for
``column pull-down'' at 3.6 and 4.5 $\micron$ and banding or ``row
pull-up" at 8 $\micron$ were applied to the calibrated individual
frames before mosaicking into a final image. Both these electronic
effects only occur at bright sources, and are strictly aligned
with the array coordinates and thus would also be minimized by
selective filtering in the Fourier domain (discussed below). The
column pull-down algorithm applied is the ``do\_pulldown''
procedure which is available on the Spitzer Science Center (SSC)
website as a contributed software tool for post-BCD (Basic
Calibrated Data) processing. The row pull-up correction used at 8
$\micron$ adjusts each row of each frame as a function of the
total flux above the 90th-percentile level of that row. The
mosaicking built into the self-calibration is an interlacing
algorithm, in which each datum ($D^i$) is averaged into only a
single sky pixel ($S^{\alpha}$). This means that the random noise
in $D^i$ translates into a completely random noise component in
$S^{\alpha}$, but it also means that for sky maps with half-size
pixels ($0.6\arcsec$) as we use here for the QSO 1700 data, the
effective coverage for a sky pixel ($\alpha$) is reduced by a
factor of 4.

{\bf Modeling Resolved Sources}

 In the full data set mosaics, the
foreground sources were identified and measured using SExtractor
(ref. [22] in main paper). The resulting source catalog is
statistically consistent (within the Poissonian uncertainties)
with the IRAC source counts (ref. [15] in main paper), both in
terms of numbers of sources and limiting magnitudes. A model image
of the foreground sky was then constructed by scaling an IRAC
point spread function (PSF) image by the intensity of each source
and adding it to an initially blank map at the appropriate
location of each source. For one version of this, we used the PSFs
that were determined during IOC and are available as FITS images
on the SSC website. However, these PSFs are only mapped within a
$\sim\ 24\arcsec\times 24\arcsec$ region, and thus neglect the
extended low-level wings of the PSFs. Thus a second model was
created in which the PSFs applied consisted of the IOC PSF cores
smoothly grafted onto the extended wings as measured by
observations of a bright (saturated) star. \footnote{Fomalhaut,
AOR ID number = 6066432} This wide grafted PSF avoids saturation
at the core while tracing the extent out to the full size of the
array ($\sim\ 300\arcsec\times 300\arcsec$). However, even with
these wide PSFs, these models cannot account for the extended
emission around many of the galaxies that are resolved by IRAC.
Therefore we developed a final model, similar to a CLEAN algorithm
(ref. [16] in main paper). To construct this model for each
channel (1) the maximum pixel intensity was located (and saved to
a list of ``clean components''), (2) the wide PSF was scaled to
half of this intensity and subtracted from the image, (3) this
process was iterated 60,000 times, saving intermediate results
every 3,000 iterations. The results were examined to identify the
appropriate iteration number where the clean components become
largely uncorrelated with the sources in the original image (i.e.
they represent randomly distributed peaks without apparent
convolution by the PSF). This limit occurs much more quickly in
the longer wavelength channels, where the relative noise level is
higher and the number of detected sources is fewer. The ``model''
is then the designated number of clean components convolved with
the wide PSF. Since this model can represent extended emission
with groups of clustered components, it is much better at removing
the extended emission of the resolved galaxies. However, even this
model will miss extended emission that is below the pixel-to-pixel
noise. Such low surface brightness features must be dealt with as
part of the analysis and interpretation of the power spectra.

{\bf Clipping individual sources}

 The original maps were clipped of
resolved sources (Galactic stars and galaxies) using an iterative
method developed for DIRBE data analysis (ref. [21] in main
paper). Each iteration calculates the standard deviation
($\sigma$) of the image, and then masks all pixels exceeding
$N_{\rm cut}\ \sigma$ along with $N_{\rm mask}\times N_{\rm mask}$
surrounding pixels. The procedure is repeated until no pixels
exceed  $N_{\rm cut}\ \sigma$. This clipping tends to miss
emission in the outer parts of extended sources and the distant
wings of very bright point sources. To help minimize this effect,
the clipping (and subsequent analysis) is followed by subtraction
of the models described above. We also clip the model images
themselves with the same algorithm at the same level of $N_{\rm
cut}$ with the mask derived from the clipping of the actual data
field. The final mask was taken to be the combined mask from the
two clipped maps, the main data and the model map. Whereas the
appearance of the residual emission around bright sources has
decreased dramatically with this extra step, the results were
practically unaffected by this extra clipping ($\lsim$ a few
percent in large scale power).

Truly robust results must be independent of the clipping
parameters to within the variations from the signal arising from
the populations of sources corresponding to the different clipping
thresholds, $N_{\rm cut}$. We explored the range for the various
clipping parameters: the results are largely independent of the
mask size and in what follows a value of $N_{\rm mask}=3$ was
adopted for displaying the final power spectra, which corresponds
to $\gsim 75$ \% pixels remaining in the clipped field at $N_{\rm
cut}=4$. For sufficiently low values of $N_{\rm cut}$ too few
pixels remained in the field for robust Fourier analysis of the
diffuse light distribution. In order to further verify robustness
of the results we also evaluated the correlation function at lower
values of $N_{\rm cut}$ when significantly fewer pixels remained.
This simple clipping algorithm proved quite efficient in
identifying the sources. Comparison with the SExtractor model
catalog shows that in channels 1 and 2 our clipping algorithm
identified over 95\% of the sources and in channel 3 and 4 the
overlap with the SExtractor catalog was nearly 100\%.

{\bf Computing power spectrum from clipped maps}

 The blanked pixels
were assigned $\delta F$=0, thereby not adding power to the
eventual power spectrum. Prior to computation of the Fourier
transform and the correlation function, we subtracted remaining
linear gradients from the maps. The Fourier transform image was
cleansed of stripes and muxbleed artifacts in the $(u, v)$
transform space by zeroing the power at frequencies corresponding
to $u = \pm n q_{mux} (n$= 1,2, 3,...) where $\pi/q_{mux} =
4\times1.2\arcsec$ which is the spacing of muxbleed artifacts.
Power was also set to zero along the $u$=0 and $v$=0 axes to
eliminate any residual signal from the gradient subtraction. The
final power spectrum of the residual diffuse light was computed by
averaging over narrow concentric rings of radius $q$. The noise
was evaluated from the $A-B$ difference maps using the same
clipping mask and weighting method as the final co-added data.

Fig. \ref{maps} shows the clipped maps for the QSO 1700 field used
in the analysis and Fig. \ref{power} plots the power spectra
deduced from the clipped maps and their noise.

\begin{figure}[h]
\renewcommand{\thefigure}{SI-\arabic{figure}}
\setcounter{figure}{0}
\begin{tabular}{cc}
 \includegraphics[width=4in]{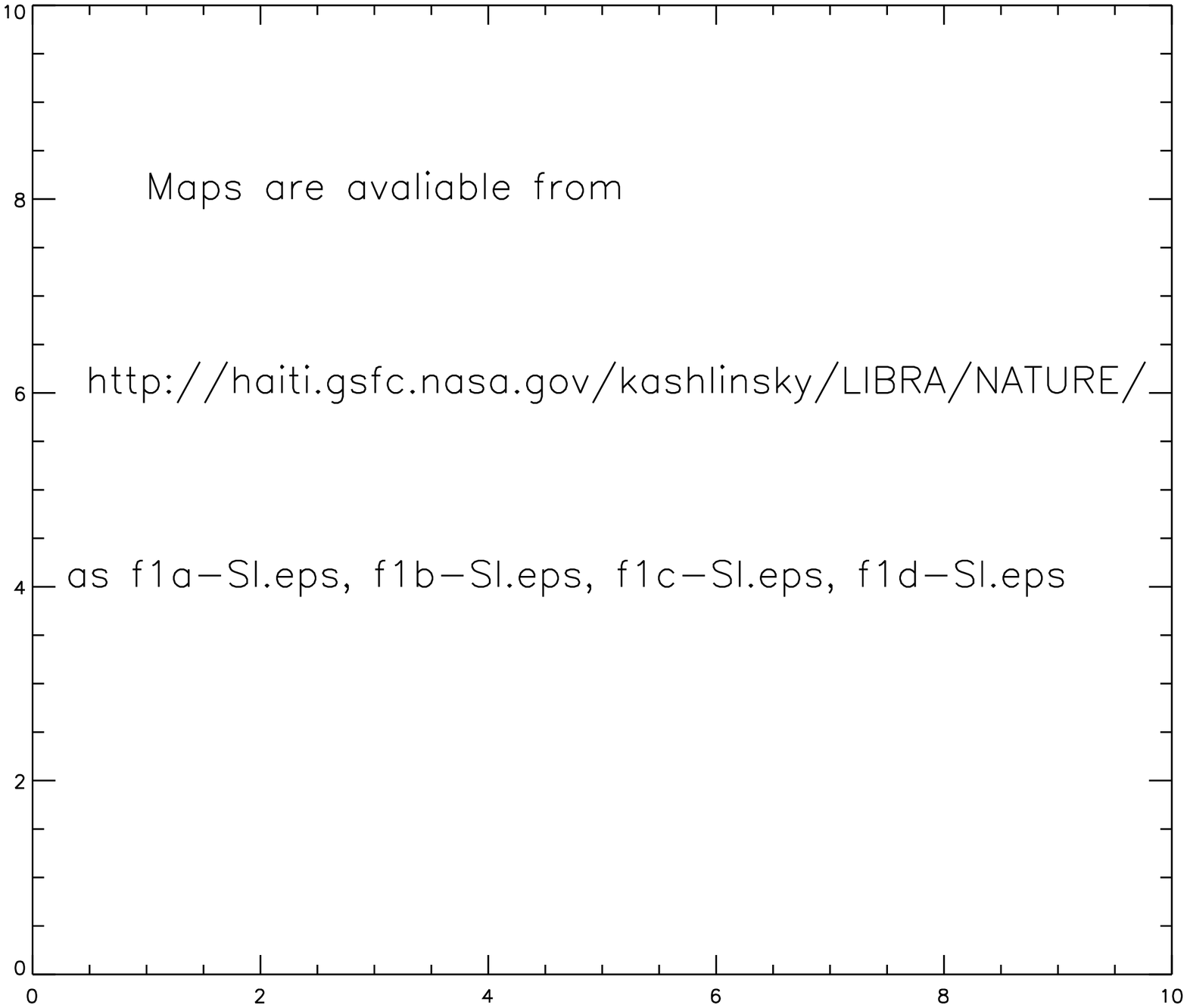} & \\
\end{tabular}
\caption{\small{Point source subtracted and clipped ($N_{\rm
cut}=4, N_{\rm mask}=3$) weighted intensity images of the QSO 1700
field for channels 1 through 4 (top to bottom). These positive
images are aligned in the horizontal direction. They are scaled
from [-2.0,4.0], [-2.2,4.4], [-9.2,18.4], and [-6.0,12] nW
m$^{-2}$ sr$^{-1}$ respectively. These ranges are proportional to
the standard deviation for each image. The regions of clipped
sources are indicated by black areas which are set to 0.0 for
further analysis.}} \label{maps}
\end{figure}

Fig. \ref{maps_cut2} shows the maps clipped at $N_{\rm cut}=2$
with $N_{\rm mask}=3$. Lowering $N_{\rm cut}$ leaves huge
multi-connected holes in the image. This precludes robustly
computing Fourier transform of the maps; hence we evaluated the
autocorrelation function, $C(\theta)=\langle \delta F(x+\theta)
\delta F(x)\rangle$ for low $N_{\rm cut}$, which is shown in Fig.
\ref{cor}. The figure shows that at every level of $N_{\rm cut}$
we find the same large scale correlations with $C(\theta)$
remaining positive to at least $\simeq 100^{\prime \prime}$. (The
slight drop in the amplitude with lower $N_{\rm cut}$ is expected
as progressively lower peaks of CIB and the instrument noise are
clipped). Note that the power spectrum, if of cosmological origin,
may depend on $N_{\rm cut}$ if the clipping pushes deep enough to
start eliminating the high-$z$ cosmological sources.
\begin{figure}[h]
\renewcommand{\thefigure}{SI-\arabic{figure}}
\setcounter{figure}{1}
\begin{tabular}{cc}
 \includegraphics[width=4in]{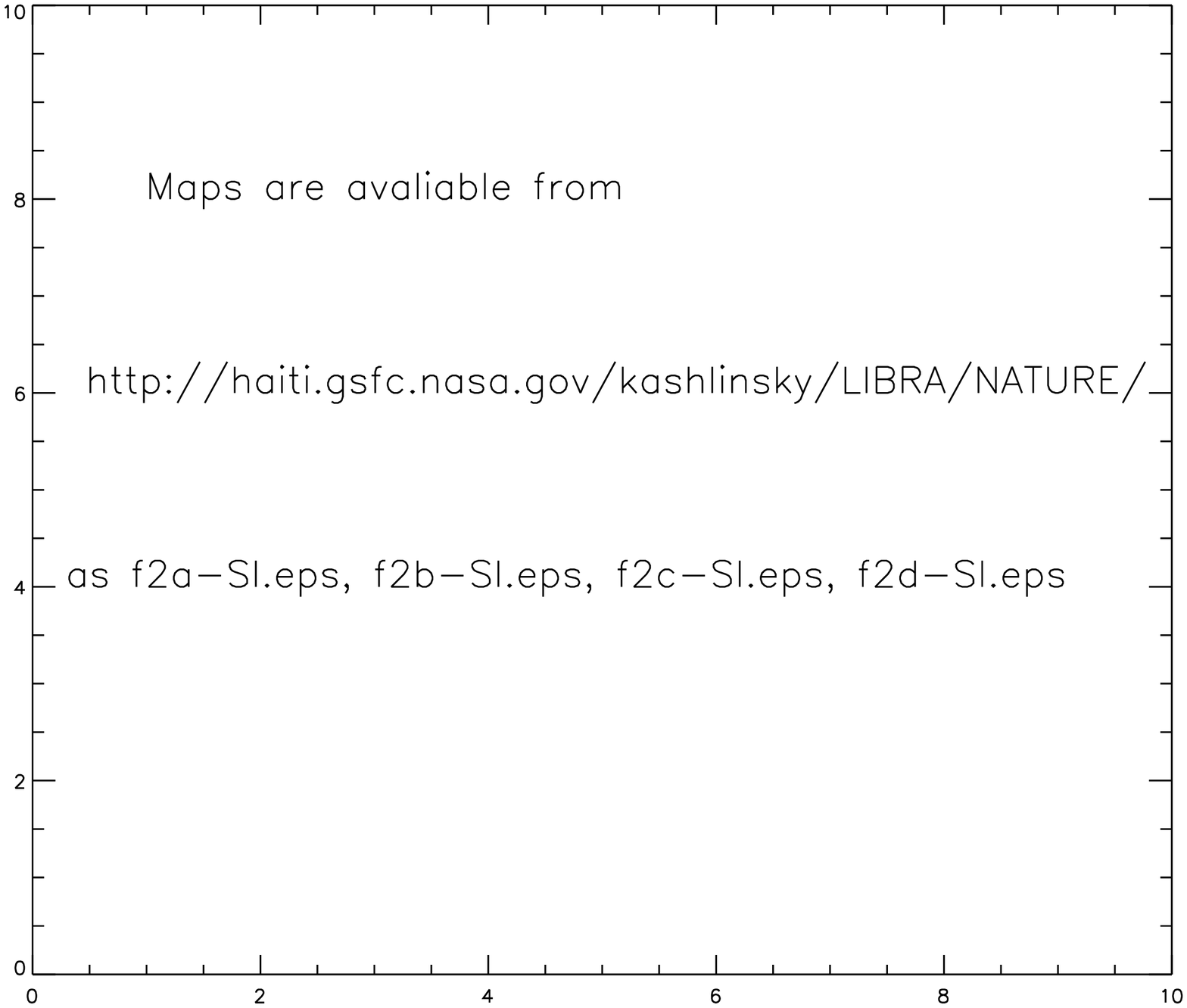} & \\
\end{tabular}
\caption{\small{Same as Fig. \ref{maps}, but for $N_{\rm
cut}=2$.}} \label{maps_cut2}
\end{figure}

\begin{figure}[h]
\renewcommand{\thefigure}{SI-\arabic{figure}}
\centering \leavevmode \epsfxsize=1.2 \columnwidth
\includegraphics[width=7in]{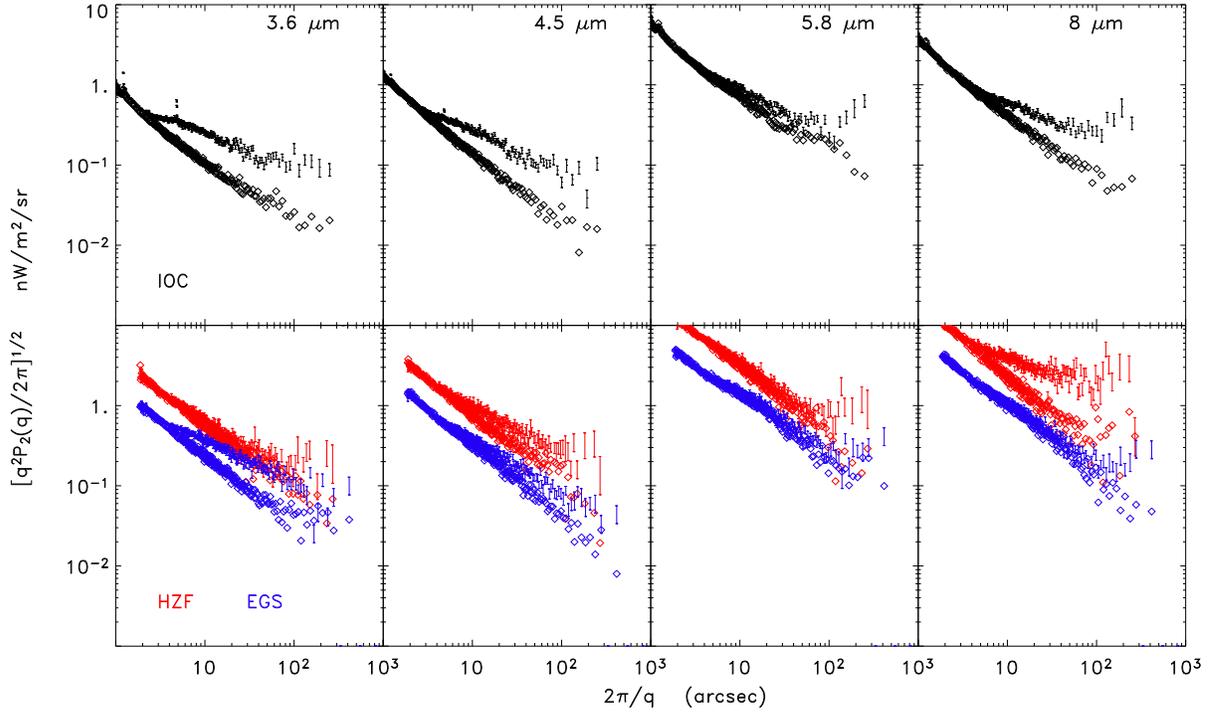} \vspace{0.5cm}
\caption[]{\small{{\bf Figure resolution reduced to conform to astro-ph reuqirements. Original figure available from http://haiti.gsfc.nasa.gov/kashlinsky/LIBRA/NATURE/}. Signal and noise from QSO 1700 field (top), the
HZF (bottom - red lines) and the EGS fields (bottom - blue lines).
The power spectrum of the residual diffuse emissions (+noise) is
plotted with its errors and the noise is shown with open diamonds.
The errors correspond to $\sqrt{N_q}$, where $N_q$ is the number
of Fourier elements that went into evaluating each value of
$P_2(q)$.} } \label{power}
\end{figure}

\begin{figure}[h]
\renewcommand{\thefigure}{SI-\arabic{figure}}
\centering \leavevmode \epsfxsize=1.2 \columnwidth
\includegraphics[width=7in]{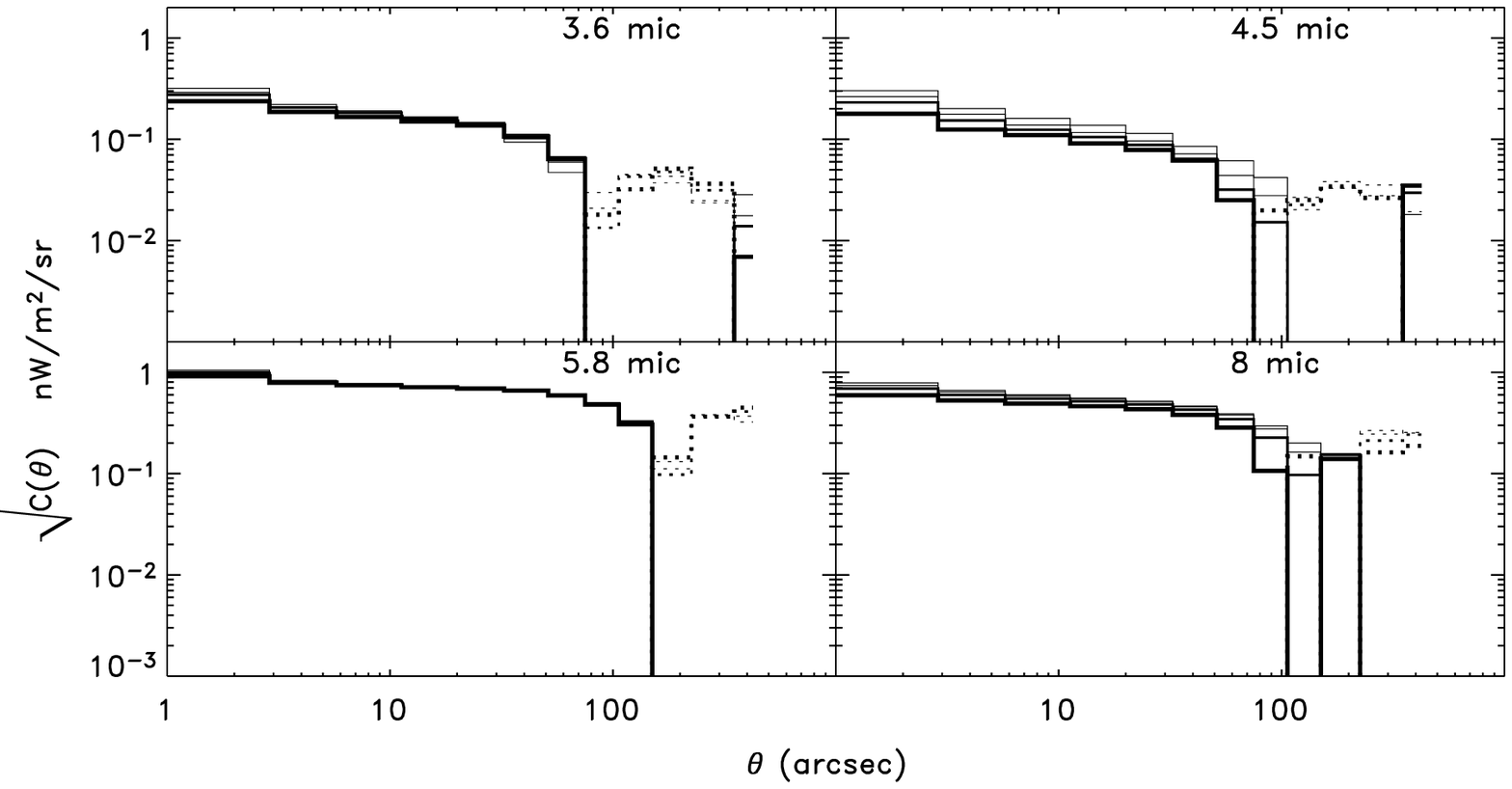} \vspace{0.5cm}
\caption[]{\small{Autocorrelation function for the QSO 1700 field
for $N_{\rm cut}=2.5,3,4$. Thick lines correspond to $N_{\rm
mask}=3$; thin lines to $N_{\rm mask}=5$. } The fraction of
clipped pixels in the QSO 1700 data at $N_{\rm cut}=4$ and $N_{\rm
mask}=3$ is between 25\% (4.5 \um) and 17 \% (5.8 \um). Dotted
lines show $\sqrt{|C(\theta)|}$ on scales where $C(\theta)<0$. }
\label{cor}
\end{figure}

{\bf Systematic instrument effects}

Systematic errors associated with the detector self-calibration,
i.e. the gain (or flat field) and the offset (or dark frame),
should produce a pattern on the sky which is the fixed pattern of
the error convolved with the dither pattern of the pointings for
each AOR. Since we do not know the pattern of the systematic
error, we created mosaics from the detector flat fields, from the
detector dark frames, and from frames consisting of a single
Gaussian point source at a fixed position on the detector.
Visually, the sort of structure that appears in these systematic
error images does not look like the actual sky images. This is
confirmed quantitatively by performing the power spectrum analysis
on these images, which in all cases show fluctuations with a
distinctly different spectrum from those of the real data. A
different test for possible instrumental effects is to check the
isotropy of the fluctuations. Instrumental effects (such as
muxbleed, column pull-down, etc.) are likely to be associated with
preferred directions on the array and on the sky as well since the
position angle of the detector was roughly constant for the
duration of the observations. As a test of isotropy for each
channel we calculated the power spectra in four 45$^\circ$-wide
azimuthal sectors of the Fourier transform's $(u,v)$ plane
centered along the lines $v = 0$, $v = u$, $u = 0$, and $v = -u$.
The sectors along the axes would be expected to be more strongly
affected by instrumental effects, but we find that there are no
large-scale variations between the power spectra in the different
sectors and only $\lsim 5\%$ variations in the overall amplitudes.
Stray light is a systematic error which remains relatively fixed
with respect to the sky rather than with respect to the detector.
There is only one star (SAO 17323, $m_V = 8.8$, $m_K = 7.6$) in
the region of the QSO 1700 field which produces visible stray
light artifacts in Channels 1 and 2 (stray light paths are
different for channels 3 and 4). However, the structure of these
artifacts are different in Channels 1 and 2, and more important
they appear in the lower coverage edges of the full images which
are excluded from the power spectrum analysis.

{\bf Interpretation of the signal with Limber equation}

More details related to this section can be found in ref. [2] of
the main paper. Whenever CIB studies encompass relatively small
parts of the sky (angular scales $\theta < 1$ radian) one can use
Cartesian formulation of the Fourier analysis. If the fluctuation
field, $\delta F(\mbox{\boldmath$x$})$, is a random variable, then
it can be described by the moments of its probability distribution
function. The first non-trivial moment is the  {projected
2-dimensional} correlation function $C(\theta) = \langle \delta
F(\mbox{\boldmath$x$}+\theta) \delta
F(\mbox{\boldmath$x$})\rangle$. The 2-dimensional power spectrum
is $P_2(q) \equiv \langle |\delta F_q|^2\rangle$, where the
average is performed over all phases. The correlation function and
the power spectrum are a pair of 2-dimensional integral
transforms, $P_2(q)= 2\pi \int_0^\infty C(\theta) J_0(q\theta)
\theta d\theta$. In the limit of small angles, the projected
2-dimensional correlation function is related to the  CIB flux
production rate, $dF/dz$, and the evolving 3-D power spectrum of
galaxy clustering, $P_3(k)$, via the Limber equation given by:
\begin{equation}
C(\theta)=\int dz {\left( \frac{dF}{dz} \right)}^2
\int_{-\infty}^\infty \xi(r;z) du \label{limber_c}
\renewcommand{\theequation}{SI-\arabic{equation}}
\end{equation}
where $r^2=c^2(dt/dz)^2 u^2 + d_A^2 \theta^2$ and $\xi(r)
=\frac{1}{2\pi^2} \int P_3(k) j_0(kr) k^2 dk$ is the two point
correlation function for 3-dimensional clustering with the power
spectrum $P_3(k)$. In the power spectrum formulation this equation
can be written as (see e.g. refs. [2,12,13,21] of the main paper
for details):
\begin{equation}
P_2(q)=\int \left(\frac{dF}{dz}\right)^2 \frac{1}{c\frac{dt}{dz}
d_A^2(z)} P_3(qd_A^{-1}; z) dz = \frac{1}{c} \int
\left(\frac{dF}{dt}\right)^2 \frac{1}{d_A^2(z)} dt
\renewcommand{\theequation}{SI-\arabic{equation}}
\label{limber_p}
\end{equation}
where $d_A(z)$ is the comoving angular diameter distance and the
integration is over the epoch of the sources contributing to the
CIB. Multiplying both sides of this equation by $q^2/2\pi$ leads,
after simple algebra, to eqs. (1),(2) in the main text of the
paper. The fluctuation of the CIB on angular scale $2 \pi/q$ can
then be expressed as:
\begin{equation}
\sqrt{\frac{q^2P_2(q)}{2\pi}}= F_{\rm CIB}
\Delta(qd_A^{-1}(\langle z \rangle))
\renewcommand{\theequation}{SI-\arabic{equation}}
\end{equation}
where the suitably averaged redshift $\langle z \rangle$ is given
by:
\begin{equation}
\Delta^2(qd_A^{-1}(\langle z \rangle)) \equiv \frac{\int dt \int
(dF/dt)^2 \Delta^2(qd_A^{-1}(z)) dt}{[\int (dF/dt) dt]^2}
\renewcommand{\theequation}{SI-\arabic{equation}}
\end{equation}
In other words,
 the fractional fluctuation on angular scale $\pi/q$ in the CIB is given by
 the average value of the r.m.s. fluctuation from spatial clustering over
 a cylinder of length $ct_0$ and diameter $\sim q^{-1}d_A^{-1}(\langle z
\rangle)$ at the effective redshift given above.

\end{document}